\documentstyle[emulateapj]{article}

\def\om{\Omega_0}
\def\lam{\lambda_0}
\def\omhat{\widehat{\Omega}_0}
\def\eqref#1{(\ref{eqn:#1})}

\def\etal{{\rm et~al. }}
\def\hmpc{\;h^{-1}{\rm Mpc}}

\def\kms{{\rm \;km\;s^{-1}}}
\def\invkms{({\rm km\;s}^{-1})^{-1}}

\def\lya{Ly$\alpha$\ }

\def\nh1{n_{\rm HI}}

\def\p1dk{P_{\rm 1D}(k)}

\newbox\grsign \setbox\grsign=\hbox{$>$} \newdimen\grdimen \grdimen=\ht\grsign
\newbox\simlessbox \newbox\simgreatbox
\setbox\simgreatbox=\hbox{\raise.5ex\hbox{$>$}\llap
     {\lower.5ex\hbox{$\sim$}}}\ht1=\grdimen\dp1=0pt
\setbox\simlessbox=\hbox{\raise.5ex\hbox{$<$}\llap
     {\lower.5ex\hbox{$\sim$}}}\ht2=\grdimen\dp2=0pt

\begin{document}

\submitted{Submitted to ApJ, September 30, 1998}
 
\title{Closing in on $\Omega_0$: The Amplitude of Mass Fluctuations 
from Galaxy Clusters and the Lyman-alpha Forest} 

\author{
David H. Weinberg$^{1}$,
Rupert A. C. Croft$^{1,2}$,
Lars Hernquist$^{2,3}$,
Neal Katz$^{4}$, 
and Max Pettini$^{5}$}
 
\footnotetext[1]{Department of Astronomy, The Ohio State University,
Columbus, OH 43210; dhw,racc@astronomy.ohio-state.edu}
\footnotetext[2]{Harvard-Smithsonian Center for Astrophysics, 
Cambridge, MA 02138; lars@cfa.harvard.edu}
\footnotetext[3]{Lick Observatory, University of California, Santa Cruz,
CA 95064}
\footnotetext[4]{Department of Physics and Astronomy, 
University of Massachusetts, Amherst, MA, 01003;
nsk@kaka.phast.umass.edu}
\footnotetext[5]{Royal Greenwich Observatory, Madingley Road, Cambridge,
CB3 OEZ, UK; pettini@ast.cam.ac.uk}

\begin{abstract} 

We estimate the value of the matter density parameter $\Omega_0$ by combining
constraints from the galaxy cluster mass function with Croft et al.'s 
recent measurement of the mass power spectrum, $P(k)$, from \lya forest data.  
The key assumption of the method is that cosmic structure formed by 
gravitational instability from Gaussian primordial fluctuations. For a 
specified value of $\Omega_0$, matching the observed cluster mass function 
then fixes the value of $\sigma_8$, the rms amplitude of mass fluctuations in 
$8\hmpc$ spheres, and it thus determines the normalization of $P(k)$ at $z=0$. 
The value of $\Omega_0$ also determines the ratio of $P(k)$ at $z=0$ to $P(k)$ 
at $z=2.5$, the central redshift of the \lya forest data; the ratio is 
different for an open universe ($\Lambda=0$) or a flat universe.  
Because the \lya forest measurement only reaches comoving scales $2\pi/k \sim 
15-20\hmpc$, the derived value of $\Omega_0$ depends on the value of the power
spectrum shape parameter $\Gamma$, which determines the relative contribution 
of larger scale modes to $\sigma_8$.  Adopting $\Gamma=0.2$, a value favored by 
galaxy clustering data, we find $\Omega_0 = 0.46^{+0.12}_{-0.10}$ for an open 
universe and $\Omega_0=0.34^{+0.13}_{-0.09}$ for a flat universe ($1\sigma$ 
errors, not including the uncertainty in cluster normalization).
Cluster-normalized models with $\Omega_0=1$ predict too low an amplitude
for $P(k)$ at $z=2.5$, while models with $\Omega_0=0.1$ predict too high
an amplitude.  The more general best fit parameter combination is 
$\om+0.2\lam \approx 0.46+ 1.3(\Gamma-0.2)$, where $\lam\equiv\Lambda/3H_0^2$.
Analysis of larger, existing samples of QSO spectra could greatly improve the
measurement of $P(k)$ from the \lya forest, allowing a determination of $\om$ 
by this method with a precision of $\sim 15\%$, limited mainly by uncertainty 
in the cluster mass function.

\end{abstract}
 
 
\section{Introduction}

In theories of structure formation based on gravitational instability
and Gaussian initial conditions, the observed mass function of galaxy
clusters constrains a combination of the
density parameter $\om$ and the amplitude of mass fluctuations.
The physics underlying this constraint is simple: massive clusters
can form either by the collapse of large volumes 
in a low density universe or by the collapse
of smaller volumes in a high density universe.  To a good approximation,
models that reproduce the observed cluster masses have 
$\sigma_8\om^{0.5} \approx 0.5$, where $\sigma_8$ is the rms mass
fluctuation in spheres of radius $8\hmpc$ and
$h \equiv H_0/(100\;\kms\;{\rm Mpc}^{-1})$.  
White, Efstathiou, \& Frenk (1993) were the first to express the
cluster normalization constraint in this form and to point out 
its insensitivity to the shape of the mass power spectrum $P(k)$.
Since $\sigma_8$ is given by an integral over 
$P(k)$, an accurate measurement of $P(k)$ could be combined with this 
constraint to determine $\om$.
Unfortunately, studies of galaxy clustering yield only the galaxy 
power spectrum, which is related to the mass power spectrum by an
unknown (or at best poorly known) ``bias factor.''

In this paper, we estimate $\om$ by combining the cluster mass
function constraint with Croft et al.'s (1998b, hereafter CWPHK)
recent determination of the linear mass power spectrum from \lya forest data.
The argument is slightly more complicated than the one just outlined
because the $P(k)$ measurement is at redshift $z=2.5$ 
and the observed units are $\kms$ rather than comoving $\hmpc$.
However, the value of $\om$ determines the linear growth factor,
which relates $P(k)$ at $z=0$ to $P(k)$ at $z=2.5$, and it determines the
relation between comoving $\hmpc$ and $\kms$ at $z=2.5$.
The scalings are different for a flat universe with a cosmological constant
($\lam \equiv \Lambda/3H_0^2 = 1-\om$) and an open, zero-$\Lambda$ universe,
so the derived $\om$ will be different in the two cases.

\section{Observational Inputs}

The $P(k)$ measurement from \lya forest data is described in detail by
CWPHK, who also test for many possible systematic uncertainties.
The key feature of this measurement, for our purposes, is the absence 
of unknown bias factors --- the method 
(introduced and extensively tested on simulations by Croft et al. [1998a])
directly estimates the linear theory
mass power spectrum, under the assumption of Gaussian
initial conditions.  The dominant uncertainty in the measurement
at present is the statistical uncertainty resulting from the
small size of the data set.  Fitting a power law,
\begin{equation}
P(k) = P_p \left(\frac{k}{k_p}\right)^n,
\label{eqn:lya}
\end{equation}
to the derived power spectrum, 
CWPHK find a logarithmic slope $n=-2.25\pm 0.18$ and a normalization
$P_p=2.21^{+0.90}_{-0.64} \times 10^7 (\kms)^3$ for the power at 
the ``pivot'' wavenumber $k_p=0.008 \invkms$, which is chosen so that errors
in $n$ and $P_p$ are statistically independent.
The measurement probes comoving scales
$2-12\hmpc$ for $\om=1$, $3-16\hmpc$ for $\om=0.5$, $\lam=0$, and
$4-22\hmpc$ for $\om=0.3$, $\lam=0.7$.

For the cluster normalization constraint, 
we adopt the results of Eke, Cole, \& Frenk (1996, hereafter ECF):
\begin{eqnarray}
\sigma_8 &= & (0.52\pm 0.04) \om^{-0.46+0.10\om} \qquad \lam=0
\label{eqn:ecf} \\
\sigma_8 &= & (0.52\pm 0.04) \om^{-0.52+0.13\om} \qquad \lam=1-\om. \nonumber 
\end{eqnarray}
These constraints are derived using the 
Press-Schechter (1974) 
formalism, cross-checked against large cosmological N-body simulations.
The 8\% uncertainty in the normalization includes a combination of
statistical uncertainties and potential systematic errors, both
observational and theoretical, as discussed by ECF.
There have been numerous other determinations of this constraint
using different methodologies and different treatments of the observational
data, and although they yield slightly different values for the normalization
and power law indices, they generally agree with equation~\eqref{ecf}
to 10\% or better (e.g., \cite{cen98}; \cite{eke98}; \cite{pen98};
\cite{viana98}).  This agreement suggests that the uncertainty
quoted by ECF is reasonable, although the various analyses share
many assumptions and often rely on the same observational data
(e.g., \cite{edge90}; \cite{henry97}).

\section{Constraints on $\om$}

Figure 1 illustrates our method, by comparing
the CWPHK measurement of $P(k)$ to the 
predictions of cluster-normalized models (i.e., models satisfying
equation~[\ref{eqn:ecf}]) that have a power spectrum
with shape parameter $\Gamma=0.2$,
in the parameterization of Efstathiou, Bond, \& White (1992, hereafter EBW).
The EBW parameterization is motivated by physical models with
scale-invariant primordial fluctuations and cold dark matter,
for which $\Gamma \approx \Omega_0 h$ if the baryon fraction is small
(see also \cite{peebles82}; \cite{bardeen86}).\footnote{The identification
of $\Gamma$ with $\om h$ is sensitive to the assumptions of a scale-invariant
inflationary spectrum and a pure CDM matter content.
For example, a CDM model with a tilted ($n<1$) inflationary spectrum would 
be roughly equivalent to an EBW model with $\Gamma<\om h$
on the scales that are relevant to our calculation.}
For our purposes, the EBW form serves as a convenient and plausible description
of the power spectrum shape that seems in reasonable accord
with observations.
From top to bottom, the curves in Figure~1 show power spectra for
$\Omega_0=0.1$, 0.2, 0.3, 0.4, 0.5, 0.6, and 1.
The left panel is for $\lam=0$, and the right for $\lam=1-\om$.
The uncertainty in the overall
normalization is indicated by the error bar on the open circle;
at the $1\sigma$ level, all points can shift coherently in amplitude
with this uncertainty.
The ($1\sigma$) error bars on the individual solid points are derived from the
dispersion among ten subsets of the \lya forest data; they represent
uncertainties in the relative amplitudes of power at different $k$.  

\begin{figure*}
\centerline{
\epsfxsize=6truein
\epsfbox[50 470 555 720]{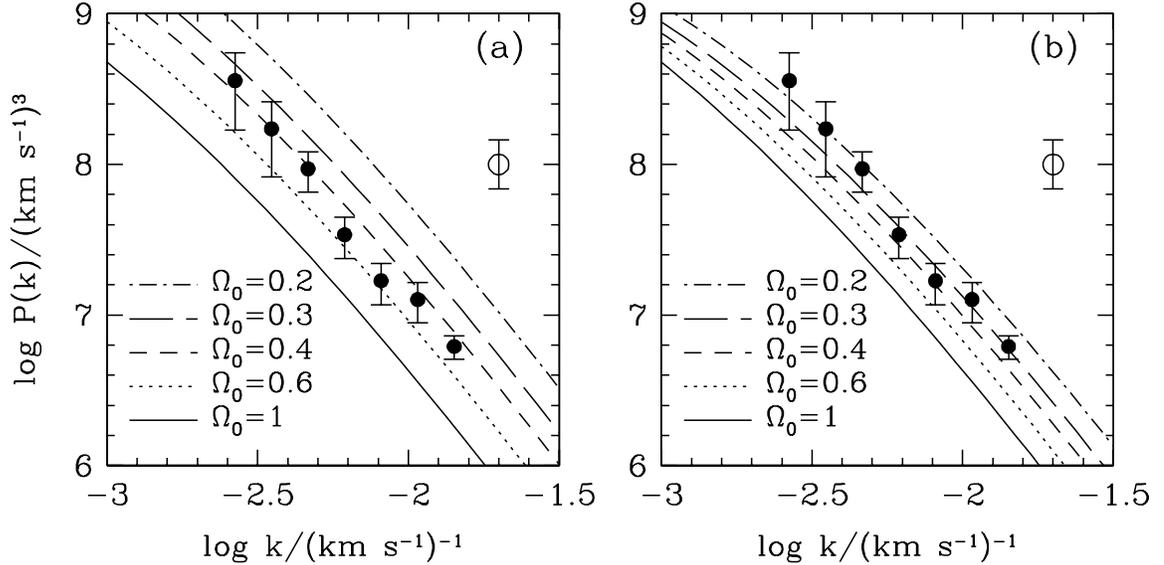}
}
\caption{
\label{fig:pk}
Filled circles (with $1\sigma$ error bars)
show the power spectrum of mass fluctuations at $z=2.5$,
derived from \lya forest spectra by CWPHK.  The error bar on the open
circle indicates the normalization uncertainty: at the $1\sigma$ level,
all points can be shifted coherently up or down by this amount.
Curves show $P(k)$ at $z=2.5$ for cluster-normalized models with 
a power spectrum shape parameter $\Gamma=0.2$ and
various values of
$\Omega_0$, as indicated.  Models with high $\Omega_0$ predict a $P(k)$
that is too low to match the \lya forest results, and models with
low $\Omega_0$ predict a $P(k)$ that is too high.
({\it a}) Open models, with $\lambda_0=0$.
({\it b}) Flat models, with $\lambda_0=1-\Omega_0$.
}
\end{figure*}

\begin{figure*}
\centerline{
\epsfxsize=6truein
\epsfbox[50 470 555 720]{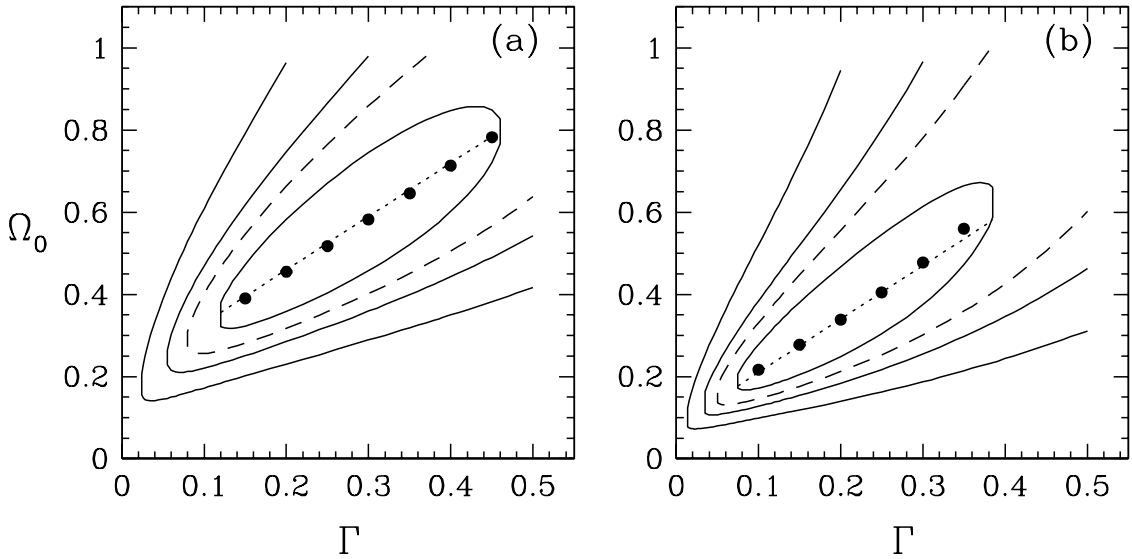}
}
\caption{
\label{fig:contour}
Constraints on the parameters of cluster-normalized power spectra,
for open models ({\it a}) and flat models ({\it b}).
Filled circles show the value of $\om$ that gives the best match
(minimum $\chi^2$) to the CWPHK power spectrum parameters at
each $\Gamma$.  Dotted lines show the ridge-line equations~\eqref{ridge}.
Solid lines show contours of $\Delta \chi^2=1$, 4, 9 in the 
$\om-\Gamma$ plane.
For a specified value of $\Gamma$, the intersection of a vertical
line with the solid contours gives the 1, 2, and $3\sigma$ confidence
intervals on $\om$.  If one ignores external information about $\Gamma$
and considers only the \lya forest data themselves, then the 
dashed contour at $\Delta\chi^2=2.30$ represents the 68\% confidence
constraint on the parameter values.
}
\end{figure*}

For $\Omega_0=1$, the normalization of $P(k)$ is low today, and reduction by
the linear growth factor $(1+2.5)^2$ between $z=2.5$ and $z=0$ puts
the predicted $P(k)$ well below the CWPHK measurement.  Conversely,
for $\Omega_0=0.1$, the $P(k)$ normalization is high at $z=0$, and
scaling back to $z=2.5$ reduces fluctuations by a smaller factor, 
yielding a predicted $P(k)$ that is too high.  
The value of $\Omega_0$ that yields the best fit to the CWPHK data is 
$\omhat=0.46$ for $\lam=0$ and $\omhat=0.34$ for $\lam=1-\om$.  
The flat model yields a lower $\om$ because of the larger linear 
growth factor between $z=2.5$ and $z=0$ in a $\Lambda$-dominated universe.  
One could also envision the $\om$ constraint by evolving the
CWPHK $P(k)$ forward in time to $z=0$: there is one and 
only one value of $\om$ for which forward extrapolation yields a
combination of $\sigma_8$ and $\om$
in agreement with the cluster constraint~\eqref{ecf}.

Because the \lya forest $P(k)$ measurement only reaches comoving scales 
$\sim 15-20\hmpc$, the value of $\om$ derived from this argument depends
on the assumed shape of $P(k)$, which determines the relative contribution of
larger scale modes to $\sigma_8$.  For Figure~1, we adopted the shape parameter
$\Gamma=0.2$, a value favored by a number of studies of large scale 
galaxy clustering (e.g., \cite{maddox90}; \cite{baugh93}, 1994; 
\cite{peacock94}; \cite{gaztanaga98}).
Figure 2 shows a more general result, confidence intervals for
cluster-normalized models in the $(\om,\Gamma)$ plane.  
We derive these by 
computing the amplitude $P_p$ and logarithmic slope $n$ of $P(k)$
at $z=2.5$ for each $(\om,\Gamma)$ combination, using
equation~\eqref{ecf} to fix $\sigma_8$, then
computing the $\Delta \chi^2$ of these $P(k)$ parameters
relative to the best fit power law parameters 
using the $\Delta\chi^2$ curves shown in figure~15 of CWPHK.
These curves (and hence the contours of Figure~2) account for 
the covariance of the individual $P(k)$ data points and 
for the contribution of uncertainty in the mean \lya forest opacity
to the uncertainty in the amplitude $P_p$.
Higher $\Gamma$ implies less large scale power and less contribution
to $\sigma_8$ from scales beyond those of the CWPHK measurement.
Since the cluster mass function determines the combination 
$\sigma_8 \om^{0.5}$, we obtain higher values of $\Gamma$
for higher values of $\om$.
The value of $\om$ affects the mapping from $\hmpc$ at $z=0$
(the units of the EBW power spectrum) to $\kms$ at
$z=2.5$, so it influences both the slope and the amplitude of
the predicted $P(k)$ --- in terms of Figure~1, changing $\om$ 
shifts the $P(k)$ curves both vertically and horizontally.
However, because the uncertainty in the slope is fairly large,
it is primarily the amplitude $P_p$ that constrains $\om$,
while the slope constrains $\Gamma$.

If we consider the \lya forest $P(k)$ and cluster normalization as
our only constraints, then the
dashed contour ($\Delta\chi^2=2.30$) represents the ``$1\sigma$''
(68\% confidence) joint constraint on $\om$ and $\Gamma$.
With the statistical uncertainty of the CWPHK measurement, we
cannot rule out the combination of high $\Gamma$ and high $\om$.  
However, if we fix the value of
$\Gamma$ based on large scale structure data (thus implicitly assuming
that biased galaxy formation does not distort the {\it shape} of
the power spectrum on large scales), then we obtain 1, 2, and $3\sigma$
constraints on $\om$ from the intersection of a vertical line with 
the solid contours, which represent $\Delta\chi^2=1$, 4, and 9.
The ridges of minimum $\Delta\chi^2$ are well described by
\begin{eqnarray}
\omhat & = & 0.46 + 1.3(\Gamma-0.2) \qquad \lam=0 ,
\label{eqn:ridge} \\
\omhat & = & 0.34 + 1.3(\Gamma-0.2) \qquad \lam=1-\om . \nonumber 
\end{eqnarray}
For $\Gamma=0.2$, the derived values of $\om$ and corresponding uncertainties
are 
\begin{eqnarray}
\om &=& 0.46~^{+0.12}_{-0.10} \;(1\sigma)~^{+0.29}_{-0.17} \;(2\sigma) 
\qquad \lam=0 , \label{eqn:omega} \\
\om &=& 0.34~^{+0.13}_{-0.09} \;(1\sigma)~^{+0.32}_{-0.16} \;(2\sigma) 
\qquad \lam=1-\om .  \nonumber 
\end{eqnarray}
The fractional uncertainty in $\om$ is smaller for open models because
the stronger $\om$-dependence of the fluctuation growth factor in an
open universe increases the sensitivity of $P_p$ to $\om$.
A critical density universe is formally ruled out at the
$3\sigma$ level for $\Gamma \la 0.2$ and at
the $2\sigma$ level for $\Gamma=0.3$.

The uncertainties in equation~\eqref{omega} do not include the
uncertainty in the cluster normalization itself.  
For a fixed value of this normalization at $z=0$,
inspection of Figure~1 shows that the
value of $P_p$ is approximately proportional to $\om^{-3/2}$ in open
models and to $\om^{-1}$ in flat models over the range $0.2 < \om < 1$.
The ECF estimate of the cluster normalization uncertainty translates 
to $\sim 15\%$ uncertainty in $\sigma_8^2$, the quantity proportional to $P_p$,
for fixed $\om$.  It therefore contributes $\sim 10\%$
(open) or $\sim 15\%$ (flat) uncertainty to our derived value of $\om$, 
to be added in quadrature to the uncertainty listed above.  
Because the uncertainty in $P_p$ with the present \lya forest sample
is large, this additional uncertainty has little effect on our error bars.

\section{Assumptions, Caveats, and Prospects}

Equation~\eqref{omega} is the principal result of this paper.  
It rests, however, on a number of assumptions:

1.
{\it Primordial fluctuations are Gaussian, as predicted by inflationary
models for their origin.}  This assumption is built into the determination
of the cluster normalization constraint and into the
normalization of the \lya forest $P(k)$.
It is this assumption that allows us to combine results from \lya forest
spectra, which respond mainly to ``typical'' ($\sim 0-2\sigma$) fluctuations
in the underlying mass distribution, with results from rich clusters,
which form from rare, high fluctuations of the density field.
The assumption is supported by studies of galaxy count
distributions (e.g., \cite{bouchet93}; \cite{gaztanaga94}; \cite{kim98}),
the topology of the galaxy distribution in redshift surveys
(e.g., \cite{gott87}; \cite{colley97}; \cite{canavezes98}),
by other large scale structure statistics (e.g., \cite{weinberg92}),
and by the statistics of cosmic microwave background (CMB) anisotropies
(\cite{colley96}; \cite{kogut96}).
However, current constraints still leave room for some non-Gaussianity
of the primordial fluctuations, perhaps enough to affect 
the $\om$ value determined by our approach.

2.
{\it The primordial power spectrum has approximately the EBW form with 
shape parameter $\Gamma \approx 0.2$.}  This assumption allows us to 
to calculate
the contribution of large scale modes to $\sigma_8$ once $P(k)$ is fixed
to match the amplitude determined from the \lya forest.
Changing $\Gamma$ changes the best fit value of $\om$
by $\Delta\omhat = 1.3(\Gamma-0.2)$,
and the influence of $\Gamma$ on the uncertainty in $\om$ can be read
from Figure~2.  A radical departure from the EBW form of $P(k)$,
such as truncation at $2\pi/k \sim 20\hmpc$, would have a more 
drastic impact on our conclusions, but it would be clearly inconsistent
with galaxy clustering data.
Gazta\~naga \& Baugh (1998, a further examination of the results in
\cite{baugh93}, 1994) argue that the EBW form of $P(k)$ predicts a
turnover that is too broad to match angular clustering in the APM galaxy
survey. However, scales near the turnover make little 
contribution to $\sigma_8$,
so our derived $\om$ would not change much if we adopted the
Gazta\~naga \& Baugh (1998) power spectrum instead of a $\Gamma=0.2$
EBW model.  

3.
{\it The cluster mass determinations used to obtain equation~\eqref{ecf} 
are correct.}  Cosmological N-body simulations give straightforward
predictions of cluster masses for specified cosmological parameters,
but cluster masses are not directly observed --- they 
are inferred with aid of assumptions
from galaxy motions, X-ray data, or gravitational lensing.  
The approximate agreement of these different methods 
(see, e.g., \cite{wu98}) supports the
robustness of these mass determinations, but the agreement is
not yet demonstrated with high precision, and the physics of cluster
formation could cause a breakdown of the standard
assumptions that would systematically
affect all three methods in the same direction.  
Many papers, including ECF, compare model predictions to
observed X-ray temperatures instead of inferred masses, but this approach 
still requires assumptions to translate theoretically predicted masses into 
cluster gas temperatures.  On the whole we regard the cluster normalization 
constraint as fairly robust, and comparisons between hydrodynamic
simulations (e.g., \cite{evrard96}; \cite{bryan98}; \cite{pen98})
and expanded X-ray temperature samples should solidify and refine it over
the next few years.  However, it is worth noting that analyses
of peculiar velocity data imply different normalizations, ranging
from $\sigma_8\om^{0.6} \approx 0.375$ (\cite{willick97}; \cite{willick98})
to $\sigma_8\om^{0.6} \approx 0.8$ (\cite{kolatt97}; \cite{zaroubi97};
\cite{freudling98}; \cite{zehavi98}).\footnote{The VELMOD studies 
(\cite{willick97}; \cite{willick98}) find
$\beta_I \equiv \om^{0.6}/b_I \approx 0.50$, which we translate to 
$\sigma_8\om^{0.6} = \sigma_{8I}(\sigma_8/\sigma_{8I})\om^{0.6} =
\sigma_{8I}\beta_I \approx 0.375$ using the measured clustering of
IRAS galaxies, which implies
$\sigma_{8I} \approx 0.75$ (\cite{fisher94}; \cite{moore94}).
This translation implicitly
assumes that the bias factor $b_I$ affecting the peculiar velocity
analysis is the same as the rms fluctuation ratio $\sigma_{8I}/\sigma_8$.
The most recent analysis using the POTENT method finds a much higher
$\beta_I=0.89$ (\cite{sigad98}), which translates to 
$\sigma_8\om^{0.6}\approx 0.67$.}
These normalizations would imply substantially different values of $\om$,
as is evident from Figure~1.

4.
{\it The physical picture of the \lya forest that 
underlies Croft et al.'s (1998a) method of $P(k)$ determination is correct.}
The essential feature of this picture is that most \lya forest absorption
arises in moderate density fluctuations ($\rho/\bar{\rho} \sim 0.1-10$)
of the diffuse, photoionized intergalactic medium, leading to a tight
relation between local mass overdensity and \lya optical depth
(\cite{croft97}; \cite{weinberg98}).  This picture is derived from
hydrodynamic cosmological simulations (\cite{cen94}; Zhang et al.\ 1995, 1998;
\cite{hernquist96}; \cite{miralda96}; \cite{wadsley97}; \cite{theuns98b}),
and a similar description was developed independently as a semi-analytic
model of the forest (\cite{bi93}; \cite{bi95}; \cite{bi97}; \cite{hui97}).
It is empirically supported by the success of the simulations and
semi-analytic models in reproducing observed properties of the \lya forest
(see the above papers and \cite{dave97}, 1998; \cite{rauch97}; 
\cite{zhang97}; \cite{bryan98a}; \cite{theuns98a}).  
It is also supported by the 
coherence of \lya absorption across widely separated lines of sight,
which gives direct evidence that the absorbing structures are low
density (\cite{bechtold94}; \cite{dinshaw94}, 1995; 
\cite{rauch95}; \cite{crotts98}).
Small scale ``cloudlet'' structure is ruled out by the
nearly perfect correlation of \lya absorption along lines of sight towards
gravitationally lensed QSOs (\cite{smette92}, 1995; \cite{rauch97a}).
The $P(k)$ determination method relies on general properties of this
physical picture, not on details of particular simulations or
a particular cosmological model.

5.
{\it The CWPHK determination of $P(k)$ is correct.}  If assumptions
(1) and (4) are correct, then the most likely source of a systematic
error larger than the estimated statistical uncertainty would be
an error in the adopted value of the mean opacity of the \lya forest.
CWPHK take this value from Press, Rybicki, \& Schneider (1993) and
incorporate Press et al.'s estimated statistical uncertainty into
the $P(k)$ normalization uncertainty.  Rauch et al.\ (1997) find a 
similar value of the mean opacity from a small sample of Keck HIRES
spectra.  However, some other determinations (\cite{zuo93}; \cite{dobrzycki96})
yield significantly lower mean opacities, and these would in turn
imply higher $P(k)$ normalizations and lower estimates
of $\om$ (see Figure~1).  CWPHK examine a number of other potential
sources of systematic error and find none that are as large as the 
statistical uncertainty, and they obtain consistent results from high
and low redshift halves of the data sample and from a second independent
data set.  Nonetheless, since this is the first determination of $P(k)$
from \lya forest data (except for the illustrative application to a
single high resolution spectrum in Croft et al.\ [1998a]),
it should be treated with some caution until it is confirmed.
At the $1\sigma$ level, the best fit values of $P_p$ and $n$
depend on the selection of the data and the parameter fitting procedure,
and the statistical error bars are themselves uncertain because they are
estimated by breaking the data into small subsets.  For our constraints
on $\om$ it is $P_p$ that matters much more than $n$, and for this
parameter we believe that the CWPHK error estimate is likely to be
conservative.

There are good prospects for checking these assumptions and improving the
precision of the $\om$ measurement with existing or easily obtainable data.
Statistical properties of high resolution spectra, such as the flux decrement
distribution function (\cite{miralda96}, 1997; \cite{rauch97}),
can independently constrain the amplitude of $P(k)$ on these scales,
and can test assumptions (1) and (4).  Larger samples of moderate
resolution spectra can provide new determinations of $P(k)$ with
smaller statistical uncertainties.  These will test assumption (5),
and because a precise determination of $n$ will tightly constrain $\Gamma$,
they will also test assumption (2), though even with a larger data set
we will probably need to extrapolate $P(k)$ to larger scales with an assumed
form in order to calculate $\sigma_8$.
With a sample of $\sim 100$ moderate resolution spectra, it should be
possible to reduce the uncertainty in the amplitude of $P(k)$ well below
the uncertainty in the cluster normalization constraint.
In this limit, the fractional uncertainty in $\om$ in flat models is
similar to the fractional uncertainty in $\sigma_8^2$ at fixed $\om$
from cluster normalization, and it is smaller by a factor of $3/2$ in
open models (see the discussion at the end of \S 2).

Our current results clearly favor a low density universe over a critical
density universe.  However, a conspiracy of small errors could still make
our results consistent with $\om=1$, without requiring a drastic violation
of any of the above assumptions.  For example, if we increase the cluster
normalization by $1\sigma$, decrease the CWPHK value of $P_p$
by $1\sigma$, and adopt $\Gamma=0.3$ instead of $\Gamma=0.2$, then
our best fit $\om$ for open models rises from 0.46 to 0.84.
Analysis of larger QSO samples should make the discrimination between
critical density and low density models more decisive in the near future.

Other recent determinations of $\om$ include the estimate 
$\om \approx 0.2$ from careful analyses of cluster mass-to-light
ratios (\cite{carlberg96}, 1997b) and estimates based on cluster
evolution that range from $\om \approx 0.3-0.5$ 
(\cite{bahcall97}; \cite{carlberg97a}; \cite{henry97};
\cite{bahcall98}; \cite{eke98})
to $\om \approx 0.7-1$ 
(\cite{blanchard98}; \cite{reichart98}; \cite{sadat98}; \cite{viana98}).
Eke et al.\ (1998) and Viana \& Liddle (1998) provide good accounts of
the current systematic uncertainties in the cluster evolution technique.
Despite its short history, we believe that the method adopted here will 
ultimately lead to a more compelling measurement of $\om$ than either
cluster mass-to-light ratios or cluster evolution, because it is
independent of complex galaxy formation physics on the one hand
and of systematic uncertainties in high redshift cluster masses on
the other.  A systematic error in cluster mass determinations at $z=0$
would affect all three methods, in the same direction.

We have not considered models with space curvature {\it and} non-zero $\lam$
in detail.
However, for $0 < \lam < 1-\om$ the best fit $\om$ should
lie between that of the open and flat cases illustrated in Figure 2.  
We have investigated results for open, non-zero $\lam$ models with
$\Gamma\approx 0.2$ and find that the best fit parameter combinations
approximately satisfy $\om+0.2\lam=0.46+1.3(\Gamma-0.2)$. 
As expected, our method is sensitive
primarily to $\om$, because of its direct influence on cluster 
normalization, and is only weakly sensitive to $\lam$.
It therefore complements measurements of cosmic acceleration
using Type Ia supernovae,
which most tightly constrain a combination that is
approximately $\om-\lam$ (see the constraint diagrams in
Kim [1998] and Riess et al.\ [1998]).
It also complements measurements of the angular location 
of the first acoustic peak in the CMB anisotropy spectrum, 
which most tightly constrain $\om+\lam$
because the peak location is sensitive to 
space curvature (\cite{kamionkowski94}).

Our result strengthens the Type Ia supernova case for a 
non-zero cosmological constant (\cite{kim98a}; \cite{riess98})
because it rules out $\lam=0$ models with very low $\om$.
It is consistent with current CMB anisotropy data
for either a flat or an open universe (see, e.g., \cite{hancock98};
\cite{lineweaver98}).
Constraints on ($\om,\lam$) parameter combinations from all three methods
should become substantially more precise in the near future.
The combination of the
three measurements should yield good, non-degenerate
determinations of $\om$ and $\lam$ and hence an empirical test of the
theoretical prejudice that favors a flat universe.
Alternatively, the three methods may yield inconsistent results,
indicating either that the assumptions underlying at least one
of the methods are incorrect or that the combination of pressureless
matter and a constant vacuum energy does not adequately describe
the energy content of our universe.

\acknowledgments 
This work was supported by NASA Astrophysical Theory Grants NAG5-3111, 
NAG5-3922, and NAG5-3820, 
by NASA Long-Term Space Astrophysics Grant NAG5-3525,
and by NSF grants AST-9802568 and ASC 93-18185.

\end{document}